\documentclass[showpacs,aps,prl,preprint,onecolumn,superscriptaddress]{revtex4}
\usepackage{graphicx}
\usepackage{dcolumn}
\usepackage{amsmath}
\usepackage{soul,xcolor}
\usepackage[normalem]{ulem}
\usepackage{mathrsfs}

\begin{document}

\title{Topological valley plasmons in twisted monolayer-double graphene moir\'{e} superlattices}

\author{Weiwei Luo}
\email{weiwei.luo@nankai.edu.cn}
\affiliation{The Key Laboratory of Weak-Light Nonlinear Photonics, Ministry of Education, School of Physics and TEDA Applied Physics Institute, Nankai University, Tianjin 300457, China}
\affiliation{Collaborative Innovation Center of Extreme Optics, Shanxi University, Taiyuan, Shanxi 030006, China}

\author{Jiang Fan}
\affiliation{The Key Laboratory of Weak-Light Nonlinear Photonics, Ministry of Education, School of Physics and TEDA Applied Physics Institute, Nankai University, Tianjin 300457, China}

\author{Alexey B. Kuzmenko}
\affiliation{Department of Quantum Matter Physics, University of Geneva, 1211 Geneva, Switzerland}

\author{Wei Cai}
\affiliation{The Key Laboratory of Weak-Light Nonlinear Photonics, Ministry of Education, School of Physics and TEDA Applied Physics Institute, Nankai University, Tianjin 300457, China}

\author{Jingjun Xu}
\email{jjxu@nankai.edu.cn}
\affiliation{The Key Laboratory of Weak-Light Nonlinear Photonics, Ministry of Education, School of Physics and TEDA Applied Physics Institute, Nankai University, Tianjin 300457, China}

\date{\today}

\begin{abstract}
In topological photonics, artificial photonic structures are constructed for realizing nontrivial unidirectional propagation of photonic information. On the other hand, moir\'{e} superlattices are emerging as an important avenue for engineering quantum materials with novel properties. In this paper, we combine these two aspects and demonstrate theoretically that moir\'{e} superlattices of small-angle twisted monolayer-bilayer graphene provide a natural platform for valley protected plasmons. Particularly, a complete plasmonic bandgap appears stemming from the distinct optical conductivities of the ABA and ABC stacked triangular domains. Moreover, the plasmonic crystals exhibit nonzero valley Chern numbers and unidirectional transport of plasmonic edge states protected from inter-valley scattering is presented.
\end{abstract}

\maketitle
 Graphene plasmons, hybrids of Dirac quasiparticles and photons, exhibit low-loss, strong electromagnetic confinement and electrical tunability\cite{koppens2011graphene,woessner2015highly,ni2018fundamental}. Graphene plasmons provide excellent opportunities for exploring light-matter interactions at the nanoscale, which is promising for applications in integrated photonics\cite{vakil2011transformation} and bio-sensing\cite{rodrigo2015mid}. Particularly, constructing graphene plasmonic crystal (GPC) provides an efficient approach for modulating plasmonic bandstructures\cite{silveiro2013plasmonic,xiong2019photonic,xiong2021programmable}. By introducing the concept of topology, unidirectional propagation of graphene plasmons protected against disorder and backscattering can be realized\cite{jin2017infrared,pan2017topologically,jung2018midinfrared}. Especially, infrared topological graphene plasmons are predicted by breaking time-reversal-symmetry with magnetic fields\cite{pan2017topologically,jin2017infrared}. Meanwhile, the valley binary degree of freedom can be utilized by breaking inversion symmetry\cite{lu2016valley,ma2016all,liu2021valley}, and topologically robust transport of valley-locked graphene plasmons were presented theoretically\cite{jung2018midinfrared}. However, the requirements of complex artificial geometries and configurations obstruct the experimental realizations of topological graphene plasmons.

On the other hand, by stacking and twisting different layers of Van der Waals materials, two-dimentional (2D) moir\'{e} superlattices are emerging as an important avenue for engineering quantum materials with novel properties\cite{morell2010flat,cao2018correlated,cao2018unconventional,shimazaki2020strongly,jin2019observation,hao2021electric,du2023moire}. In particular, in small-angle twisted heterostructures, atomic reconstruction effects generate domain walls separating two kinds of periodically arranged domains with different stacking orders\cite{yoo2019atomic,weston2020atomic,halbertal2022unconventional}, with the period $a$ that can be flexibly manipulated by varying twist angle $\theta$ according to the formula $a=a_0/[\mathrm{2sin}(\theta/2)]$ ($a_0$ is the lattice constant of individual layer). Specifically, $a$ can reach hundreds of nanometers for twisted graphene layers with $\theta<0.1^\circ$\cite{sunku2018photonic,zhang2022domino}, comparable to the wavelength of graphene plasmons. Consequently, the regular modifications of the electronic structures and optical properties provide a natural and lithography-free host for GPCs, as demonstrated from recent studies on plasmonic crystal response of  small-angle twisted bilayer graphene moir\'{e} superlattices\cite{sunku2018photonic,brey2020plasmonic}. However, the stacking domains which dominant most regions of the superlattices hold identical plasmonic response, and the limited discrepancy near the domain walls does not open a complete plasmonic bandgap.

\begin{figure}
\centerline{\includegraphics[width=11cm]{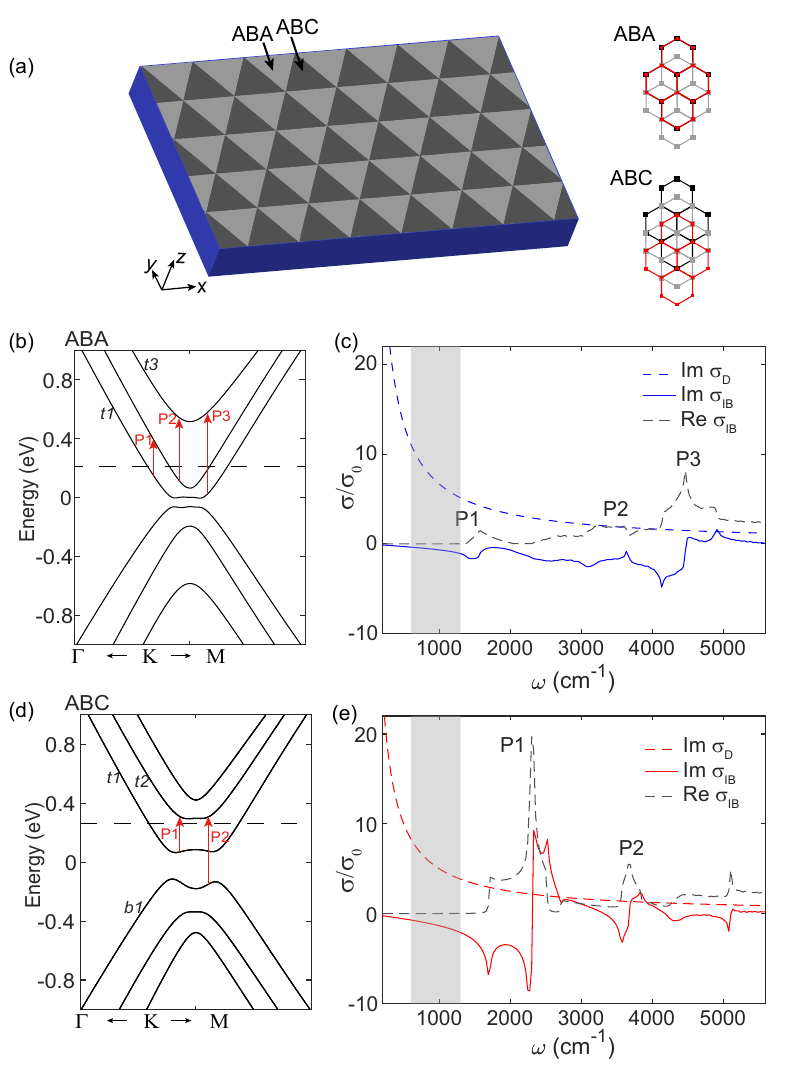}}
\caption{Moir\'{e} superlattices made of twisted monolayer-bilayer graphene. (a) Sketch of a moir\'{e}  superlattice made of domains of ABA and ABC stacked graphene. The right part represents atomic structures of the two stacking orders, where the symbols with black, gray and red colors label the first, second and third atomic layers, respectively. (b) Calculated electronic bandstructure of the ABA graphene. (c) Frequency dependence of the calculated optical conductivities for the ABA graphene, including imaginary part of the intraband transition term (Im$\sigma_\textrm{D}$), imaginary (Im$\sigma_{\textrm{IB}}$) and real (Re$\sigma_{\textrm{IB}}$) parts of the interband transition term. (d,e) Same as (b,c), but for the ABC graphene.
}
\label{fig1}
\end{figure}

With the rapid explorations of twisted heterostructures, reconstructed moir\'{e} superlattices with two types of stacking domains of different optical responses are emerging\cite{halbertal2021moire,zhang2022domino,halbertal2022unconventional,rosenberger2020twist,moore2021nanoscale}. In this work, the effect of triangular domains with distinct optical conductivities on the plasmonic bandstructures is revealed. In particular, plasmon properties of small-angle twisted monolayer-bilayer graphene (tMBG) is investigated, whose moir\'{e} superlattices consist of triangular domains with the Bernal (ABA) and the rhombohedral (ABC) stacking\cite{li2020global,zhang2022domino}, as illustrated in Fig.\ref{fig1}a. The ABA and ABC graphene have different electronic bandstructures\cite{mak2010electronic,bao2011stacking,morpurgo2015abc} where the ABA graphene is a semi-metal with a tunable band overlap, while the ABC one is a semiconductor with a gate-tunable band gap and a flat band. Therefore the two stacking show distinct optical conductivities\cite{lui2011observation,luan2022imaging}. Here we demonstrate theoretically that tMBG moir\'{e} superlattice provides a natural platform for GPC, where complete plasmonic bandgap occurs. Specifically, the pronounced tail of interband transitions from the ABC graphene yields sufficient difference between plasmon response of the two stacked domains. Furthermore, the effects of nontrivial chiral valley topology of the GPC are emphasized. Finally, robust transport of graphene plasmon waves with suppressed inter-valley scattering is shown at the interfaces separating two GPCs with opposite valley Chern numbers. Our study motivates further explorations of novel photonic phenomena in the rich platform of reconstructed moir\'{e} superlattices.

Generally, plasmon wave vector $q$ of graphene is related to its conductivity $\sigma(\omega)$ through\cite{koppens2011graphene}
 \begin{equation}\label{eq1}
q=2\omega\varepsilon_0\varepsilon_r i/\sigma(\omega),
 \end{equation}
where $\omega$ is the light frequency and $\varepsilon_r$ is the effective dielectric constant of environment. To reveal plasmon response of the ABA and ABC graphene under doping, their electronic bandstructures were calculated from the tight-binding model self-consistently\cite{avetisyan2009electric} and optical conductivities can be obtained from the Kubo formula for the intraband ($\sigma_{\mathrm{D}}$) and interband ($\sigma_{\mathrm{IB}}$) terms (supplementary note 1), following the approach adopted in previous studies\cite{lui2011observation,ubrig2012infrared,mei2018terahertz,luan2022imaging}. Fig.\ref{fig1}b-e show the representative results for a back-doped carrier density of 2$\times$10$^{13}$ cm$^{-2}$ which is easily achievable in experiments\cite{choi2021high} (for results at other doping levels, see supplementary note 2).

While the Drude terms are similar between the two stacking orders, with the one for the ABA graphene being slightly stronger\cite{luan2022imaging} (Fig.\ref{fig1}c and e), the obtained interband terms $\sigma_{\mathrm{IB}}$ are distinct, in agreement with previous reports\cite{lui2011observation}. For the ABC graphene, the doping induced an electronic bandgap between bands $b1$ and $t1$ (Fig.\ref{fig1}d).  Consequently, the two strong transition peaks P1 and P2 in the curve of Re$\sigma_{\mathrm{IB}}$ in Fig.\ref{fig1}e correspond to transitions from band $t1$ to $t2$ and from band $b1$ to $t2$, respectively, and their energy difference reflects the size of the electronic bandgap\cite{lui2011observation}. On the other hand, the interband transitions are strongest at P3 for the ABA graphene(Fig.\ref{fig1}c)\cite{lui2011observation,ubrig2012infrared}, which is at around $\sqrt{2}\gamma_1$ and barely moves with doping ($\gamma_1$ is the nearest-neighbour interlayer coupling strength in the tight-binding model, see supplementary note 1 and 2).

\begin{figure}
\centerline{\includegraphics[width=16cm]{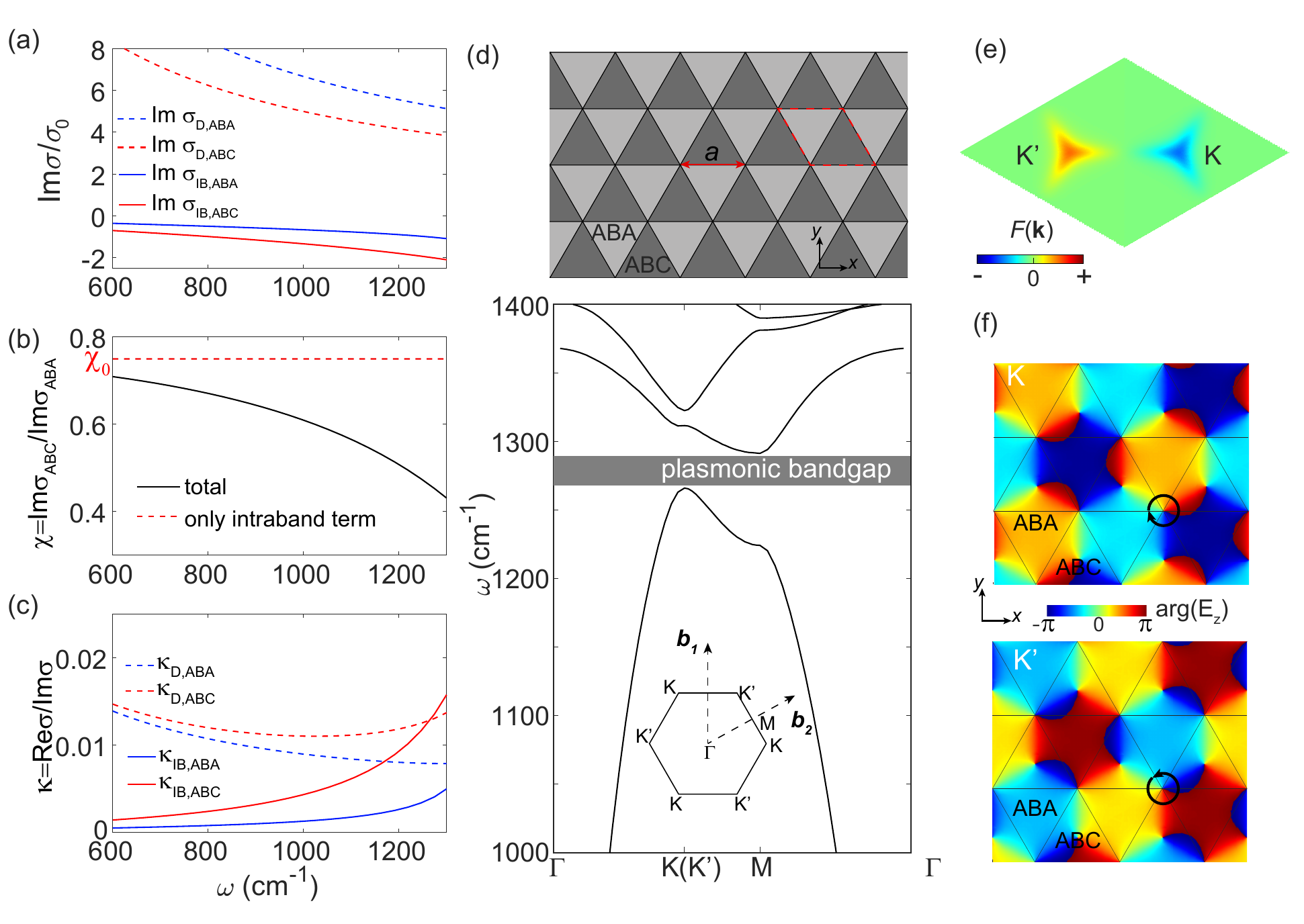}}
\caption{
Plasmonic bandstructures of the moir\'{e} superlattices. (a) Extracted curves of Im$\sigma_{\textrm{D}}$ and Im$\sigma_{\textrm{IB}}$ for the two stacked graphene within the frequency range marked in Fig.\ref{fig1}c and e (gray regions). (b) Frequency dependence of $\chi$. $\chi_0$ is the value considering only the Drude terms. (c) plots intraband ($\kappa_\textrm{D}$) and interband ($\kappa_{\textrm{IB}}$) damping factors for the two stacking orders. (d) Top: sketch of the graphene plasmonic crystal made of moir\'{e} superlattices. The dashed rhombus labels a unit cell, with lattice constant of $a$. Bottom: calculated plasmonic bandstructures of the GPC for $a$=150~nm. The inset shows the Brillouin zone and its reciprocal vectors $\textbf{b}_1$, $\textbf{b}_2$. (e) Distributions of the Berry curvature $F(\textbf{k})$ for the first plasmonic band. (f) Phase distributions of the z-components of electric fields for the K and K' valleys at the first band.
}
\label{fig2}
\end{figure}

According to Eq.\ref{eq1}, the ratio between real parts of $q$, i.e. $\mathrm{Re}(q)$, of the two stacking can be obtained from $\chi$=$\mathrm{Im}\sigma_{\mathrm{ABC}}$/$\mathrm{Im}\sigma_{\mathrm{ABA}}$. Fig.\ref{fig2}a compares the extracted Im$\sigma$  within a frequency range between 600 and 1300~cm$^{-1}$, where tails of the intraband and interband transitions are observed. If only the intraband terms are considered, a constant value of $\chi$=$\chi_0$=0.78 is obtained for all the frequencies, as shown in Fig.\ref{fig2}b. Actually, the value of $\chi_0$ depends on the Fermi velocities for the two stacking orders, and varies slightly with different doping levels ( supplementary note 2). In reality, the intraband terms dominate the conductivities at lower frequencies, and $\chi$ is close to $\chi_0$. With the increase of the light frequency, the intraband terms decrease while the interband terms contribute more. Moreover, the ABC graphene presents lower values of Im$\sigma_{\mathrm{IB}}$, stemming from the influence of the intense peak P1 shown in Fig.\ref{fig1}e. Consequently, $\chi$ decreases dramatically with the increase of the light frequency, reaching 0.5 at around $\omega$ =1200~cm$^{-1}$. Therefore, a big difference of Re($q$) between the two stacking graphene can be obtained.

Generally, graphene plasmons are weakly damped if the light frequency is far below the interband transitions and intraband term dominates\cite{koppens2011graphene,woessner2015highly}. Here, to clarify influence of the interband transitions on plasmon damping, the interband damping factor $\kappa_{\mathrm{IB}}$=Re$\sigma_{\mathrm{IB}}$/Im$\sigma$\cite{woessner2015highly,ni2018fundamental} is studied and compared with the intraband one $\kappa_{\mathrm{D}}$=Re$\sigma_{\mathrm{IB}}$/Im$\sigma$. As shown in Fig.\ref{fig2}c, with the increase of the light frequency, $\kappa_{\mathrm{IB}}$ increases with the rate faster for the ABC graphene than for the ABA one, stemming from the stronger interband transition of the ABC graphene. Nonetheless, $\kappa_{\mathrm{IB}}$ stays at a same level of $\kappa_{\mathrm{D}}$ for the considered frequency range if assuming same values of electronic broadening factors in the Kubo formula (supplementary note 1), which are both 1~meV in this calculation (a value achievable at low temperatures\cite{ni2018fundamental}). Therefore, interband transitons related plasmon damping is still quite weak, as the light frequency is well below the interband transition peaks. Similar results can be obtained for other doped carrier densities (supplementary note 2).

Have shown a large difference between plasmon response of the ABA and ABC graphene, next we calculate plasmonic bandstructures of the moir\'{e} superlattices. As illustrated in Fig.\ref{fig2}d, the interlaced triangular lattices of ABA and ABC graphene forms a natural photonic crystal for graphene plasmons, with a rhombic unit cell made of two triangular domains. The plane-wave expansion method (PEM) is employed for calculating plasmonic bandstructures, where an eigenproblem for the electromagnetic potential $\varphi(\mathbf{r})$ is solved (supplementary note 3). Fig.\ref{fig2}d presents the plasmonic bandstructures for superlattice constant $a$=150~nm. The first plasmonic band has a maximum at the K (K') point, while the second band is lowest at the M point. Particularly, a complete plasmonic bandgap of around 20~cm$^{-1}$ wide is observed at around $\omega$=1270~cm$^{-1}$. Here, domain walls between the ABA and ABC graphene are ignored for simplicity, and their influence will be discussed later.

Importantly, triangular domains with different optical conductivities break the inversion symmetry, which is the key to valley photonics\cite{lu2016valley,ma2016all,jung2018midinfrared}. Therefore, nontrivial topology of the GPC is investigated by calculating the valley Chern numbers $C_{\nu}=\frac{1}{2\pi i}\int_{\bigtriangleup_{\nu}}d^2\textbf{k} F(\textbf{k})$\cite{jung2018midinfrared}, where $\nu$ is the valley index (K or K'), $F(\textbf{k})=\nabla_\textbf{k}\times\langle\varphi_\textbf{k}|\nabla_\textbf{k}|\varphi_\textbf{k}\rangle$ is the Berry curvature, $\varphi_\textbf{k}$ is the eigenstate of $\varphi(\mathbf{r})$ at wave vector $\textbf{k}$, and $\bigtriangleup_{v}$ represents the integral triangle of each valley $\nu$. Fig.\ref{fig2}e presents the distribution of $F(\textbf{k})$ for the first plasmonic band. As expected\cite{ma2016all,liu2021valley,jung2018midinfrared}, opposite Berry curvatures are observed for the two valleys, yielding nonzero and opposite valley Chern numbers $C_{\textrm{K}}$ and $C_{\textrm{K'}}$. Besides, the broken inversion symmetry would lift of the degeneracy between the two sublattice pseudospins, and valley chiral states would appear\cite{liu2016pseudospin,lu2016valley}. This is confirmed from phase distributions of the z-components of electric fields $E_z$ in Fig.\ref{fig2}f (calculated using the finite element method). For the K(K') valley, the $E_z$ phase increases clockwise(counterclockwise) by 2$\pi$ at each unit cell corner. Thus the two valleys have opposite circular orbital angular momentums (OAMs)\cite{chen2017valley}, and unidirectional excitation of these valley chiral states can be realized by sources carrying OAM with proper chirality\cite{lu2016valley,chen2017valley}.

\begin{figure}
\centerline{\includegraphics[width=16cm]{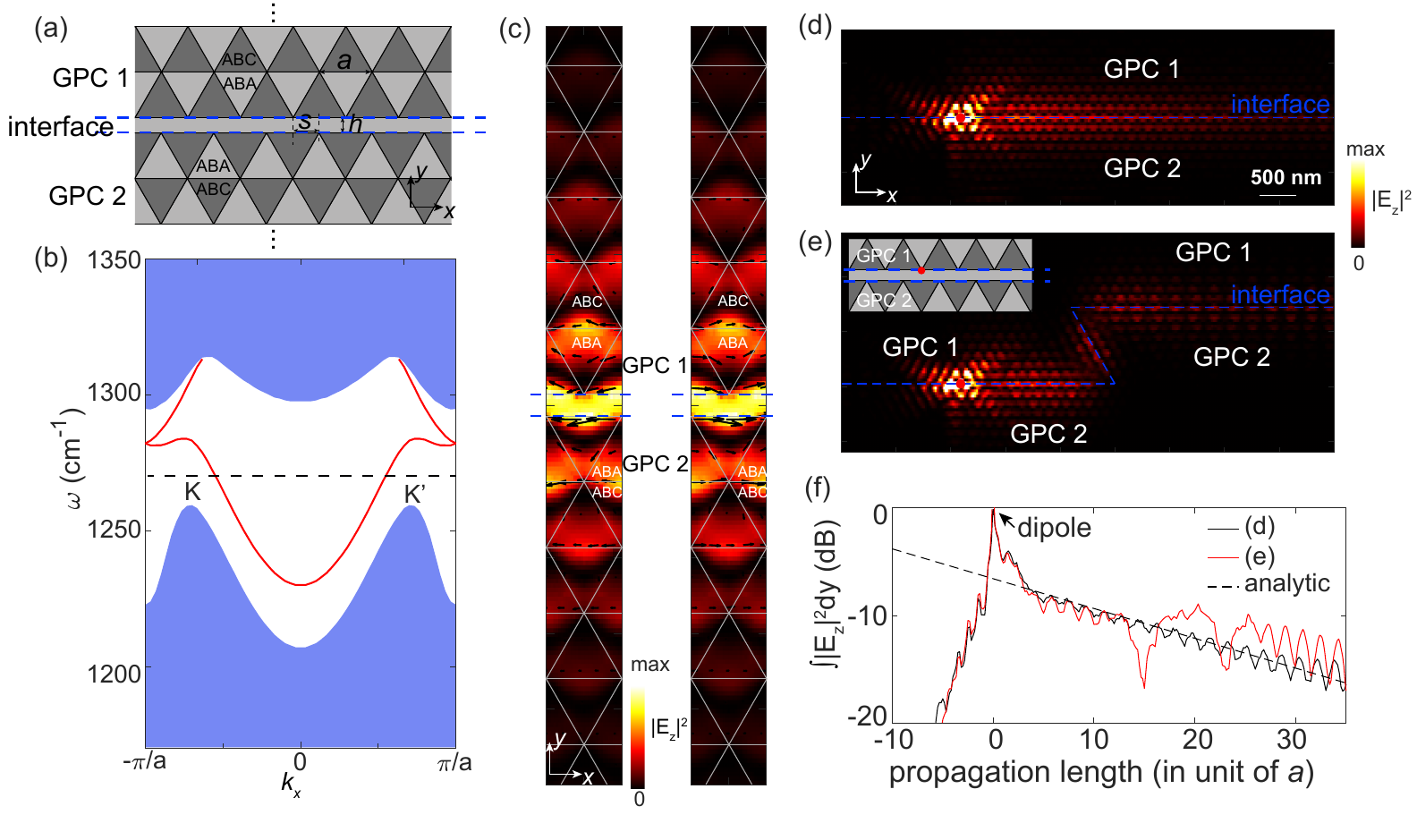}}
\caption{Valley topologically protected edge states. (a) Sketch of the studied interface structures, where an interface (marked within the blue lines, in $x$ direction) separates two plasmonic crystals (GPC1 and GPC2) with opposite valley Chern numbers. The two GPCs are shifted in $x$-direction by $s$, and gapped in $y$ direction by $h$. The interface is assumed as ABA stacked graphene. (b) Dispersion of the edge states (represented by the red curves) for wave vectors in $x$ direction ($k_y$=0). Here, the geometric parameters are $h=a/(2\sqrt{3})$, and $s=0.5a$. The blue regions mark the 2D plasmon states. (c) Spatial distributions of $|E_z|^2$ and energy flux at the edge states near the K (left) and K' (right) valleys at $\omega_0$=1270~cm$^{-1}$ as labeled in (b) (dashed line). Length and direction of the black arrows represent magnitude and direction of the energy flux, respectively. These values are extracted at 20~nm above graphene. (d,e) Propagation of graphene plasmons along the straight (d) and Z-shape (e) interfaces (labeled by blue lines) at the light frequency of 1270~cm$^{-1}$. The inset in (e) illustrates top view of the point sources (red dots), which sit 10~nm above the graphene surfaces. (f) Propagating length dependence of the plasmon powers calculated by integrating $|E_z|^2$ along the directions perpendicular to the interfaces. The analytic curve (in dashed black) is presented with a slope of $\kappa\omega_0/2v_g$.
}
\label{fig3}
\end{figure}

Topologically protected edge states can be created at the interface between two graphene plasmonic crystals with opposite valley Chern numbers\cite{jung2018midinfrared}. As an example, we study the interface structures shown in Fig.\ref{fig3}a.  The regions GPC1 and GPC2 possess domains of ABA and ABC with reversed orders, and therefore carry opposite valley Chern numbers. As plotted in Fig.\ref{fig3}b, the edge states cross the plasmonic bandgap, with opposite directions of group velocity near the two valleys. Dispersions of edge states for other geometric parameters are presented in supplementary note 4. Fig.\ref{fig3}c exhibits the local electric field confinement near the interface and the opposite directions of energy flux for the two edge states near the K (left) and K' (right) valleys. Moreover, robust propagation of graphene plasmons protected from intervalley scattering are demonstrated at $\omega_0$=1270~cm$^{-1} $in Fig.\ref{fig3}d-f. As a representative example, a Z-shape waveguide is constructed (Fig.\ref{fig3}e) and the propagation of graphene plasmons is compared with the straight one (Fig.\ref{fig3}d). Graphene plasmon waves are excited by a right-hand circularly polarized magnetic point dipole (inset of Fig.\ref{fig3}e). Besides, to model the absorption loss, a weak damping factor value $\kappa$=Re$\sigma$/Im$\sigma$=1/300 is assumed for all the graphene nanostructures, which might be achieved at low temperature for encapsulated graphene\cite{ni2018fundamental}. For both cases, the excited plasmon waves propagate along the interfaces unidirectionally. Moreover, according to the attenuation of plasmon power plotted in Fig.\ref{fig3}f, the absorption (analytic curve of slope $\kappa\omega_0/2v_g$, where $v_g$ is the group velocity) is the only source of propagation loss. Strikingly, the plasmon waves can go around the 120$^\circ$ corners without intervalley scattering. Notably, various approaches might be explored to compensate the plasmon loss, for instance, via gain media\cite{chakraborty2016gain}, stimulated electron tunneling\cite{PhysRevB.94.115301}, nonlinear optical effects\cite{you2020four} and synthesized complex frequency excitation\cite{guan2024compensating}.

In practice, fabrications of the periodic moir\'{e} superlattices are advancing with the rapid understanding of morphology\cite{yoo2019atomic,weston2020atomic,kazmierczak2021strain,halbertal2022unconventional} and improvements of stacking techniques\cite{lau2022reproducibility}. Particularly, moir\'{e} patterns with tunable periodicity and ultralow disorder have been reported recently\cite{kapfer2023programming}. Furthermore, based on the controllable switching and local arrestments of stacking domain orders in tMBG\cite{zhang2022domino}, a possible route for preparing the interface structures could be envisioned (Fig.s7 of the supplementary). Moreover, various strategies can be employed to enlarge the difference of plasmon response between the two domains, like engineering the dielectric environment\cite{karalis2005surface} and exploring the stacking domain-dependent surface functionalization\cite{hsieh2023domain}, which would increase the plasmonic bandgap and thus minimize the influence of fabrication inaccuracy. Experimentally, the edge states can be excited near a resonant antenna by far-field illumination\cite{xiong2019photonic} (Fig.s8 of the supplementary) and detected via the well-demonstrated technique of scattering-type scanning near-field microscopy (s-SNOM).

\begin{figure}
\centerline{\includegraphics[width=11cm]{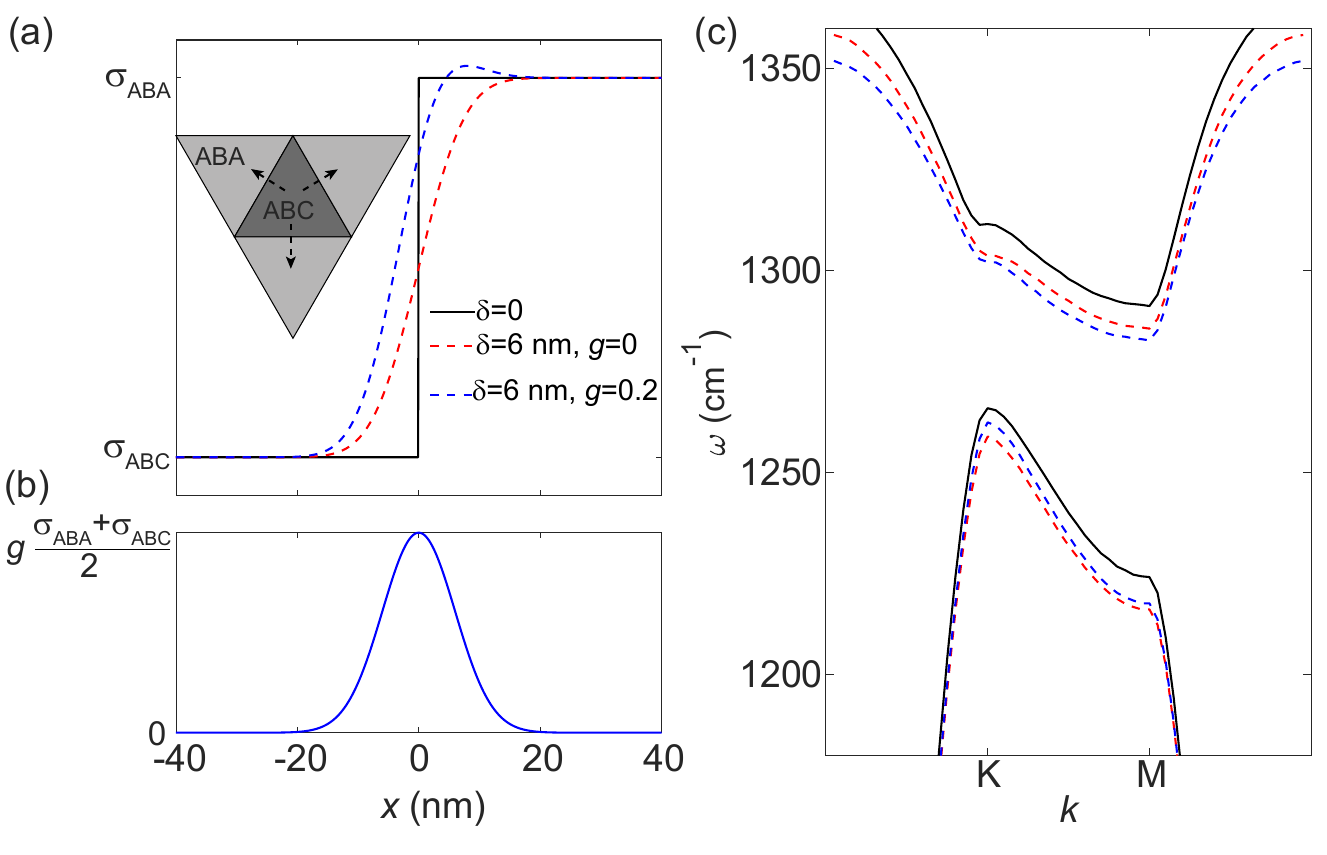}}
\caption{Influence of the domain walls on plasmonic bandstructures. (a) Conductivity profiles for the three cases: sharp boundary ($\delta$=0), transition boundary ($\delta$=6~nm, $g$=0) and transition boundary ($\delta$=6~nm, $g$=0.2). (b) Profile of the conductivity with a Gaussian distributions, centering at the middle of the boundary. (c) Bulk plasmonic dispersions calculated for the three cases shown in (a).
}
\label{fig4}
\end{figure}

Finally, influence of the domain walls is discussed. Various kinds of domain walls between the ABA and ABC graphene can exist\cite{yankowitz2014electric,zhang2022domino}. Here, for simplicity, we assume identical conductivity profile along the three directions across the domain walls as indicated in Fig.\ref{fig4}a, and consider two kinds of contributions from the walls. Firstly, instead of a infinite sharp boundary ($\delta=0$), a transition boundary with finite width $\delta$=6~nm\cite{yankowitz2014electric,zhang2022domino} is assumed, which can be described by
\begin{equation}
M(x)=\frac{\sigma_{\mathrm{ABA}}-\sigma_{\mathrm{ABC}}}{2}\textrm{erf}(\frac{x}{\sqrt{2}\delta})+\frac{\sigma_{\mathrm{ABA}}+\sigma_{\mathrm{ABC}}}{2},
\end{equation}
where $\textrm{erf}(x)$ is the error function. Secondly, a peak in the conductivity curve might emerge at the boundary stemming from the distinct electronic bandstructures like the case of bilayer graphene\cite{jiang2017plasmon}. This is described by a Gaussian function $g\frac{\sigma_{\mathrm{ABA}}+\sigma_{\mathrm{ABC}}}{2}\texttt{exp}(\frac{-x^2}{2\delta^2})$ (Fig.\ref{fig4}b), where $g$ represents weight of this term. Fig.\ref{fig4}a plots conductivity profiles for the three cases: infinite sharp boundary ($\delta$=0), transition boundary with $\delta$=6~nm, $g$=0 and transition boundary with $\delta$=6~nm, $g$=0.2. The plasmonic bandstructures for the GPC are re-calcluated, and presented in Fig.\ref{fig4}c. As observed, the transition boundaries lower the entire plasmonic bands, while the bandgap width is barely changed. Besides, although the bandgap width is reduced by including the Gaussian term, complete plasmonic bandgap is still observable for the moderate value of $g$=0.2 (close to the value for domain walls of twisted bilayer graphene in reality\cite{sunku2018photonic}).

In conclusion, small-angle tMBG moir\'{e} superlattices are demonstrated as natural GPCs with complete plasmonic bandgap.  Moreover, the inversion symmetry is broken, and valley topology of the GPC is revealed. Our studies thus provide a new avenue for realizing unidirectional propagating of graphene plasmons protected from inter-valley scattering. Therefore, various tunable and compact valley plasmonic devices including resonators, modulators and switches can be foreseen, which are promising for applications in integrated photonics and biosensing. More importantly, our studies demonstrate topological plasmon polariton effects of domains with distinct optical response, which can be generally applied for exploring various kinds of photonic phenomena with the advance of new and complex moir\'{e} superlattices\cite{vizner2021interfacial,yasuda2021stacking,hao2021electric,zhang2022promotion,zeng2023exciton,cai2023moir}, such as topological phonon polaritons\cite{guddala2021topological} in twisted hBN layers\cite{moore2021nanoscale}, topological exciton polaritons\cite{liu2020generation} in twisted transition metal dichalcogenide materials\cite{weston2020atomic,rosenberger2020twist}.

~\\
\begin{acknowledgments}
This work has been supported by the National Key Research and Development Program of China (2023YFA1407200, 2022YFA1404800), the National Natural Science Foundation of China (12004196,12127803,12074200), Guangdong Major Project of Basic and Applied Basic Research (2020B0301030009), Changjiang Scholars and Innovative Research Team in University (IRT13\_R29) and the 111 Project (B23045). The work of A. B. K. was supported by the Swiss National Science Foundation.

\end{acknowledgments}


\newpage

\end{document}